\documentclass[final,5p,times,twocolumn]{elsarticle}
\usepackage{amssymb,color}

\newcommand{\jcap}{J. Cosmol. Astropart. Phys.}

\newcommand{\mnras}{Mon. Not. R. Astron. Soc.}

\newcommand{\prd}{Phys. Rev. D}
\newcommand{\prl}{Phys. Rev. Lett.}

\usepackage{hyperref}

\begin{document}

\begin{frontmatter}

\title{Revisiting the analysis of axion-like particles with the Fermi-LAT gamma-ray observation of NGC1275}

\author{Ji-Gui Cheng$^1$,Ya-Jun He$^1$,Yun-Feng Liang$^{1,*}$,Rui-Jing Lu$^1$ and En-Wei Liang$^{1,\dagger}$}

\address{$^{1}$ Guangxi Key Laboratory for Relativistic Astrophysics, School of Physics Science and Technology, Guangxi University, Nanning 530004, China \\
liang-yf@foxmail.com, lew@gxu.edu.cn}

\begin{abstract}
In this work, we re-analyze the Fermi-LAT observation of NGC 1275 to search for axion-like particle (ALP) effects and constrain ALP parameters. Instead of fitting the observed spectrum with ALP models, we adopt an alternative method for the analysis of this source which calculates the irregularity of the spectrum. 
With the newly used method, we find no spectral oscillation for the NGC 1275 and rule out couplings $g_{a\gamma}>3\times10^{-12}\,{\rm GeV^{-1}}$ around ALP mass of $m_a\sim$ 1 neV at 95\% confidence level, which is more stringent than the previous results. We also show that the constraints can be further improved by combining the observation of PKS 2155-304. We suggest that with more sources taken into account, we could obtain a much wider exclusion region. 
\end{abstract}

\begin{keyword}
Dark matter \sep Gamma rays: general \sep Axion-like particle
\end{keyword}

\end{frontmatter}

\section{Introduction}
\label{intro}
Historically, to solve the strong CP problem, Roberto Peccei and Helen Quinn proposed a new gauge symmetry \citep{Peccei1977}. The spontaneously broken of this symmetry results in a new particle, axion. 
Axion-like particles (ALPs) have properties similar to those of axion. One major difference between them is that both the mass $m_a$ and the coupling strength to photons $g_{a\gamma}$ of axion are proportional to the energy scale \citep{Peccei1977, WeinbergS1978, WilczekF1978}, while they are independent parameters for ALPs \citep{JaeckelJ2010}. Although ALPs cannot account for the strong CP problem, both axions and ALPs could be good candidates to interpret the mysterious dark matter (DM) existed in our universe \citep{AnselmAA1982,Abbott&Sikivie1983,Dine&Fischler1983,PreskillJ1983,CicoliM2012, DiasAG2014,MarshDJE2017}. Thanks to the improved instrument sensitivity and abundant observation data accumulated in recent years, astronomical investigations on ALPs can be performed with gamma-ray observations. Considering the gamma-ray observations are generally sensitive only to the ALP parameter space, below we focus mainly on ALPs.

The coupling of ALPs to photons is described by a Lagrangian term of the form $\mathcal{L}_{a\gamma}=g_{a\gamma}\vec{E}\cdot\vec{B}a$, where $\vec{E}$, $\vec{B}$ and $a$ represent electric, magnetic, and ALP fields, respectively. This term leads to a property of ALPs that they can convert into photons and vice versa in an external electromagnetic field (Primakoff effect \cite{RaffeltG1988}). Such a feature provides people a channel to detect ALPs indirectly by examining the behavior of photon beams after passing through magnetic fields. Via the Primakoff process, many laboratory experiments have been designed to search for signatures of the photon-ALPs conversion, such as ALPS, CAST and ADMX and so on \citep{EhretK2010, ArikM2011, AsztalosSJ2010}. Based on the same effect, astrophysical observations could be used to study ALPs as well. Gamma-ray emissions from distant astrophysical sources would pass through a series of magnetic field environments before they reaching the observer on the Earth. Their spectra will be modulated if ALP fields exist and unique characteristics appear. More explicitly, the conversion between photons and ALPs will lead to oscillation or absorption structures in the observed spectrum, while the signals predicted by a standard astrophysical radiation mechanism usually have continuous spectrum. A wide range of the parameter space of ALP model has been probed according to observations from various target sources \citep{MirizziA2009,BelikovAV2011,TavecchioF2012,WoutersD2012,deAngelisA2013,Meyer2013,MeyerM2014a,ReesmanR2014,WoutersD2014,BerenjiB2016,Kohri&Kodama2017,VogelH2017,  ZhangC2018,DayF2018,GalantiG2018,LiangYF2019,GalantiG2020,Libanov2020,LongGB2020,cta_alp,BiXJ2021,LiHJ2021,GuoJG2021,Yuan:2020xui}

The photon-ALP oscillation can also account for the problem of low observed opacity of the Universe for TeV photons. Very high energy (VHE) photons travelling in the space will interact with the extragalactic background light (EBL) and transform to electron/positron pairs, which prevent them from propagating a long distance. Therefore, the Universe is opaque to sufficiently distant TeV sources. However, previous researches showed that we can see VHE gamma-ray photons in the regime of optical depth $\tau>2$ even adopting a minimum EBL model \citep{deAngelisA2009, DominguezA2011, HornsD2012a, RubtsovGI2014}. This issue can be easily addressed if we take into account an ALP effect \citep{deAngelisA2011, HornsD2012b, MeyerM2014b}. The VHE photons emitted from distant sources may convert into ALPs, which do not interact with EBL and propagate unimpededly in the universe, so that they can travel much a longer distance. These ALPs convert back into gamma-ray photons in the Milky Way's magnetic fields, making them detectable by the observer on the Earth.

Currently, studies on searching for ALPs or constraining the ALP properties using gamma-ray observations have been conducted over a wide variety of astrophysical environments. With the HESS observations of the distant BL Lac object PKS 2155-304 (z=0.116), HESS collaboration quantified the degree of irregularity of the spectrum and constrained the ALP parameters. As they didn't find any signal, more stringent upper limits on the $g_{a\gamma}$ can be provided for $m_a$ between 15 and 60 neV comparing to laboratory experiments \citep{Abramowski2013}. Making use of 6 years of Fermi-LAT observations, Fermi-LAT collaboration searched for irregularities in the gamma-ray spectrum of the radio galaxy NGC 1275. Though no indication of ALP effect has been found, the parameter space of $g_{a\gamma}>5\times10^{-11}\;{\rm GeV}^{-1}$ for ALP masses $0.5-5\;\mathrm{neV}$ was ruled out according to the null result \citep{AjelloM2016}. In addition, analyses of some Milky-Way sources (pulsars and supernova remnants) show intriguing indications of the spectral oscillation \citep{MajumdarJ2018, XiaZQ2018}. However, the best-fit parameters are in tension with the limit from the CAST helioscope and further analysis suggested that the 
spectrum irregularity in SNR may be due to systemic effects\citep{XiaZQ2019} (see however \cite{Choi2020} and \cite{AdamanePallathadkaG2020} for possible solution to the tension between CAST limit and the positive hint for ALPs). Meanwhile, the absence of counterpart gamma rays from a Milky-Way supernova explosion may probe the parameter space where ALPs could constitute the entire DM in the Universe \citep{MeyerM2017}.

In this paper, we re-analyze the Fermi-LAT observations of NGC 1275 to further investigate the ALP properties. Since previous work is based on six years of Fermi-LAT data \citep{AjelloM2016}, the exposure time of the source at present has been doubled. Moreover, in recent years, the gamma-ray emission of this source is in a flaring state with higher fluxes (see Fig. \ref{pic:ngc1275_lc}). We speculate the higher statistics of the data can provide more stringent constraints on the ALP parameters \footnote{This is one of 
the motivations that prompt us to start the work. However, at the stage of we preparing our paper, we note that another work have reported the updated ALP results of NGC 1275 \cite{AdamanePallathadkaG2020}. Nevertheless, we use different analysis method and concern on different topic.}. 
In addition, a different analysis method is used in our work. 
Rather than fit the source’s spectrum with both null and ALP models and compare the goodness of fit between the two, we instead calculate the degree of irregularity of the spectrum. 
The method previously has been adopted in the ALP analyses of HESS data \cite{Abramowski2013} and X-ray observations of galaxy clusters \cite{Wouters&Brun2013}.

\section{ALP propagation in magnetic fields}
\label{sec:propa}
Our calculation of the ALP propagation refers mainly to \cite{MeyerM2014a} and the references therein.

The ALP-photon coupling is described by the Lagrangian \citep{ZhangC2018}
\begin{eqnarray}
\mathcal{L}=-\frac{1}{4} F_{\mu \nu} F^{\mu \nu}+\frac{1}{2}\left(\partial_{\mu} a \partial^{\mu} a-m_{a}^{2} a^{2}\right)-\frac{1}{4} g_{a \gamma} a F_{\mu \nu} \tilde{F}^{\mu \nu} \\ \nonumber
+\frac{\alpha^{2}}{90 m_{e}^{4}}\left[\left(F_{\mu \nu} F^{\mu \nu}\right)^{2}+\frac{7}{4}\left(F_{\mu \nu} \tilde{F}^{\mu \nu}\right)^{2}\right]
\end{eqnarray}
For an initially polarized photon beam propagating through a single homogeneous magnetic field domain, the propagation equation is
\begin{equation}
\label{eq:propa}
    \left(i\frac{\mathrm{d}}{{\mathrm{d}x}_{3}}+E+\mathcal{M}_0\right)\Psi(x_{3})=0
\end{equation}
with 
\begin{equation}
    \Psi(x_{3})=\left(A_{1}(x_{3}),A_{2}(x_{3}),a(x_{3})\right)^{T}
\end{equation}
where $A_{1}(x_{3})$ and $A_{2}(x_{3})$ are the polarization states along $x_{1}$ and $x_{2}$ axis respectively, $a(x_{3})$ denotes the ALP state. $\mathcal{M}_0$ is the photon-ALP mixing matrix
\begin{equation}
    \mathcal{M}_0 = \left(
        \begin{array}{ccc}
        \Delta_{\perp}-\frac{i}{2}\Gamma & 0 & 0 \\
        0 & \Delta_{\parallel}-\frac{i}{2}\Gamma & \Delta_{a\gamma} \\
        0 & \Delta_{a\gamma} & \Delta_{a}
        \end{array}
    \right)
\end{equation}
The terms $\Delta_{\perp}$, $\Delta_{\parallel}$, $\Delta_{a}$, $\Delta_{a\gamma}$ depend on $m_a$, $g_{a\gamma}$, photon energy $E$, strength of the transverse magnetic field $B_{\rm T}$ and electron density $n_{\rm e}$. The absorption rate $\Gamma$ is included in the mixing matrix to take into account the EBL absorption. We use the EBL model \citep{DominguezA2011} to calculate the absorption rate.

For an initially unpolarized photon beam, it is convenient to work with the density matrix 
\begin{equation}
    \rho(x_{3})=\Psi(x_{3})\Psi(x_{3})^{\dagger}
\end{equation}
which obeys the von-Neumann-like commutator equation
\begin{equation}
    i\frac{\mathrm{d}\rho}{\mathrm{d}x_{3}}=\left[\rho,\mathcal{M}_0\right]
\end{equation}
The final state of the photon-ALP beam is given by the solution to the above equation
\begin{equation}
    \rho(x_{3})=\mathcal{T}(E)\rho(0)\mathcal{T}^{\dagger}(E)
\end{equation}
with $\mathcal{T}$ the full transfer matrix which can be determined by solving Eq.(\ref{eq:propa}). The explicit solution can be found in \cite{BassanN2010}. The photon survival probability in the final state is thus given by
\begin{equation}
\label{eq:survial_prob}
    P_{\gamma\gamma}=\mathrm{Tr}((\rho_{11}+\rho_{22})\mathcal{T}(E)\rho(0)\mathcal{T}^{\dagger}(E))
\end{equation}
where $\rho_{11}={\rm diag}(1,0,0)$, $\rho_{22}={\rm diag}(0,1,0)$ {and the initial state $\rho(0)=1/2{\rm diag}(1,1,0)$}.

From source to the observer, gamma rays pass through four main magnetic field environments: the fields in the jet (JET), the intracluster magnetic field (ICMF), the intergalactic magnetic field (IGMF), and the Galactic magnetic field of the Milky Way (GMF). For the ALP parameters and the Fermi-LAT energy range considered in this work, the JET and IGMF components are negligible.

The typical field strength of ICMF is in the range of 1 - 10 $\mu{\rm G}$. Faraday Rotation Measure (RM) observations and magnetohydrodynamic simulations show that the magnetic field is turbulent, which is usually modeled as a divergence-free homogeneous isotropic field with Gaussian turbulence with zero mean and a variance $\delta_{B}$. The power spectrum follows a power-law $M(k)\propto k^q$ with the wave number $k_{L}<k<k_{H}$ ($k_{L}=2\pi/\Lambda_{\rm max}$, $k_{H}=2\pi/\Lambda_{\rm min}$). The radial profile of the magnetic field strength is $B_{r}=B_{0}\left[n_{\rm e}(r)/n_{\rm e}(0)\right]^{\eta}$. For the profile of the electron density the Eq.(4) in \cite{ChurazovE2003} is used. To better compare with the previous work (i.e. \citet{AjelloM2016}), we use the following parameters to model the ICMF component, $r_{\rm max}=500\,{\rm kpc}$, $\delta_B=10\,{\rm \mu G}$, $\eta=0.5$, $\Lambda_{\rm min}=0.7\,{\rm kpc}$, $\Lambda_{\rm max}=35\,{\rm kpc}$, $q=-2.8$. The uncertainty to the ALP effect induced by the choice of the values of these parameters have been discussed in \cite{AjelloM2016, MeyerM2014b, ZhangC2018}.

The GMF consist of a large-scale regular component and a small-scale random component. Since the coherence length of the turbulent component is much smaller than the photon-ALP oscillation length, we do not include it in the calculation. For the regular magnetic field, we take into account the model developed by \cite{JanssonR2012} that has been widely adopted in similar researches \citep{AjelloM2016, MeyerM2017, MajumdarJ2018,LiangYF2019}.

\section{Fermi-LAT data analysis and calculation of the spectrum irregularity}
\label{sec:data}
The Fermi-LAT is a pair-conversion gamma-ray telescope sensitive to $>$30 MeV photons \cite{atwood2009}. In this work, we analyze 12 years of Fermi-LAT Pass 8 data towards the direction of NGC 1275, which range from August 4, 2008 to August 4, 2020 (239557417 - 616478947 in MET). Considering the relatively bad performance of LAT below 100 MeV and low statistics above 500 GeV, the data are extracted in the energy range of 100 MeV to 500 GeV with {\tt SOURCE} event class (evclass=128, evtype=3). We adopt an region of interest (ROI) of $15^\circ$ centered on the radio position of the target source, NGC 1275. To eliminate the contamination from the Earth Limb, events with zenith angles larger than $90^\circ$ are ignored in the analysis ($Z_{\rm max}=90^\circ$). A data quality cut {\tt (DATA\_QUAL>0)\&\&(LAT\_CONFIC==1)} is applied to guarantee that the data are suitable for scientific use. The instrument response functions (IRFs) of {\tt P8R3\_SOURCE\_V2} is chosen to match the extracted data. To fit the observation data, both the target source and background sources are included in the model. For background sources, we consider all the 4FGL point sources $25^\circ$ around the NGC 1275 together with the Galactic and extragalactic diffuse components (described by {\tt gll\_iem\_v07.fits} and {\tt iso\_P8R3\_SOURCE\_V2\_v1.txt} respectively). The script \textit{make4FGLxml.py}\footnote{https://fermi.gsfc.nasa.gov/ssc/data/analysis/user} is employed to generate \textit{xml} model file.

\begin{figure}
    \centering
    \includegraphics[width=0.46\textwidth]{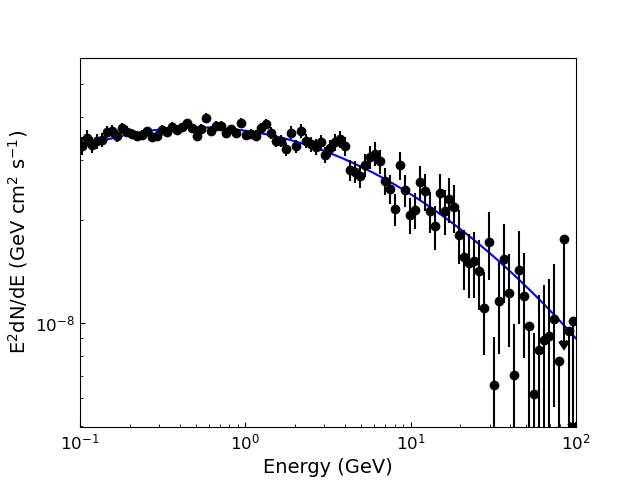}
    \caption{The SED of NGC 1275. It contains 100 energy bins. In each bin, a maximum likelihood analysis is performed to determine the flux and error bar. The model used to fit the overall spectrum is a {\tt LogParabola} function. Upper limits are plotted if TS$<25$.}
    \label{pic:ngc1275_sed}
\end{figure}

\begin{figure*}
    \centering
    \includegraphics[width=0.9\textwidth]{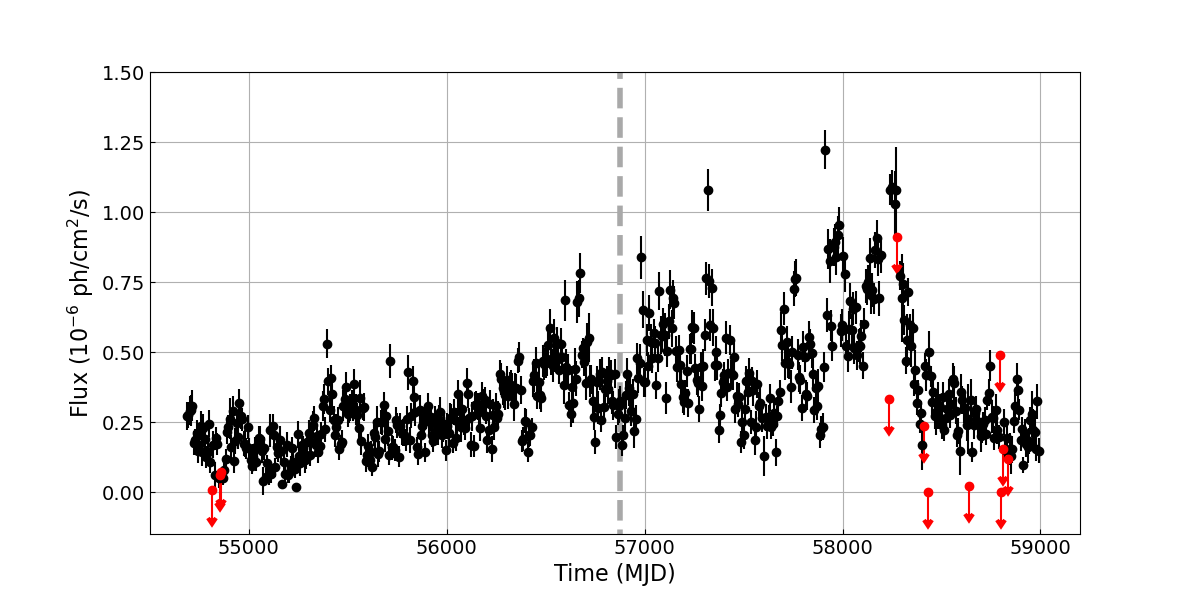}
    \caption{The light curve of NGC 1275. A 7-day bin size is adopted and the likelihood analysis is performed in each time bin. The vertical grey line marks the time 6 years since the Fermi launch. We can see that the source is in flaring states in recent years.}
    \label{pic:ngc1275_lc}
\end{figure*}

The likelihood analysis are implemented by the command {\texttt{gtlike}}. 
We first perform a likelihood analysis using the whole data set covering all the energy range and time range (we call it global fit) to obtain a good estimation of the background parameters. During the global fit, the spectral parameters of the point sources within $5^\circ$ and normalization parameters of all the point sources within ROI and the two diffuse components are left free to vary, and energy dispersion is enabled to achieve better accuracy. Next, we divide the data set into multiple energy and time bins to derive the spectral energy distribution (SED) and light curve (LC) of NGC 1275. For SED we use 100 bins (0.1 - 100 GeV) and for LC we use a weekly bin size. In each bin, a likelihood analysis is performed to derive the corresponding flux. All spectral indices are fixed in this procedure. If the fit is subject to a converge problem \footnote{In this case, the resulted error bar of the flux will be extremely small, so that the calculated $I$ value is incorrect.}, for background point sources we will only keep the prefactor of a 0.5$^\circ$ nearby source free to vary. The resulting SED and LC are demonstrated in Fig. \ref{pic:ngc1275_sed} and Fig. \ref{pic:ngc1275_lc}. As shown in Fig. \ref{pic:ngc1275_sed}, a {\tt LogParabola} model 
\begin{equation}
\label{eq:logparabola}
    \frac{\mathrm{d}N}{\mathrm{d}E}=N_0\left(\frac{E}{E_b}\right)^{-(\alpha+\beta\mathrm{log}(E/E_b))},
\end{equation} 
can give a good overall estimation of the source spectrum.

As mentioned above, in this work, we do not directly fit the observed spectrum with ALP model. Instead, we calculate the degree of irregularity ($I$) of the spectrum.
The idea is, the more significant the ALP effect is, the higher degree of irregularity will appear in the spectrum. If a set of ALP parameters ($m_a$, $g_{a\gamma}$) predicts a high value of $I$, but the one derived from actual observation ($I_{\rm obs}$) is very low, then the parameters are disfavoured.
In our analysis, considering the random nature of the turbulent magnetic fields and the statistical fluctuation, we will simulate 50,000 (see below) energy spectra for each pair of ALP parameters ($m_a$, $g_{a\gamma}$) to derive the expected distribution of $I$.
The distribution then is compared to the $I_{\rm obs}$ to determine whether the ALP parameters are ruled out or not at the 95\% confidence level (C.L.).
Hence, how to quantify the spectral irregularity is crucial in our analysis. 

In \cite{Abramowski2013}, the local power-law behavior is tested over three consecutive energy bins (triplet) of the spectrum, the residuals of the middle bins from the power law are quadratically summed to form the irregularity estimator (see \cite{Abramowski2013} for details). However, this method cannot be applied directly to the analysis here. The small bin size of the NGC 1275 SED makes the spectral structure induced by ALP effect (after energy dispersion) is much wider than the bin size. Therefore each triplet behave always as a power-law and the irregularity cannot be evaluated correctly. Alternatively, we propose to define the level of irregularity as 
\begin{equation}
\mathcal{I}=\sum_{i}\sum_j \frac{\left(\phi^{\rm pl}_{i,j}-\phi_{i,j}\right)^{2}}{\sigma_{i,j}^2}.
\label{eq:m1}
\end{equation}
The spectrum is fitted in a series of energy windows of (0.5$E_{\rm mid}$, 1.5$E_{\rm mid}$). In each window, a power law function is used to fit the spectrum. The narrow window width ensure the power law approximation is reasonable. The $E_{\rm mid}$ is the mid energy of each window with increments in steps of 0.5$\sigma_{E}$ starting from 200 MeV. Similar method called `sliding window analysis' is often used in line-like signal searches \cite{liang16line,fermi15line}. The $j$ and $i$ denote the $j$th bin in the $i$th window. The $\phi^{\rm pl}$ is the best-fit PL spectrum, and $\phi$, $\sigma$ are the measured flux and corresponding uncertainty, respectively.

The irregularity can also be estimated with the scatter of the fit residuals, as suggested in Ref.\cite{WoutersD2012},
\begin{equation}
    I=\frac{1}d\sum_k^{N}\frac{(\phi_{\rm w/oALP}(\vec{\theta})-\phi_{k})^2}{\sigma_{k}^2}=\chi^2/d
\label{eq:m2}
\end{equation}
where $d$ is the degree of freedom and $\vec{\theta}$ denotes all the spectral parameters of best-fit null model. One can notice that the $I$ defined here is in fact a reduced-$\chi^2$ of the best fit. The $N=100$ is the bin number of the SED.
A shortage of this method is that an intrinsic overall spectrum $\phi_{\rm w/oALP}$ need to be assumed first. However for NGC 1275, as we have shown in Figure \ref{pic:ngc1275_sed}, a Log-Parabola function can give a good overall estimate of the global spectrum (see also the $I$ value derived in Sec. \ref{sec:result}), thus would not introduce too much additional bias. For simplicity, we use Eq.(\ref{eq:m2}) to quantify the level of irregularity in this work. For the upper limit bins (i.e. TS$<25$), we ignore them in calculating the $I$ (Eq.(\ref{eq:m2})), the corresponding bins will also be ignored in the below simulations.

\begin{figure}
    \centering
    \includegraphics[width=0.46\textwidth]{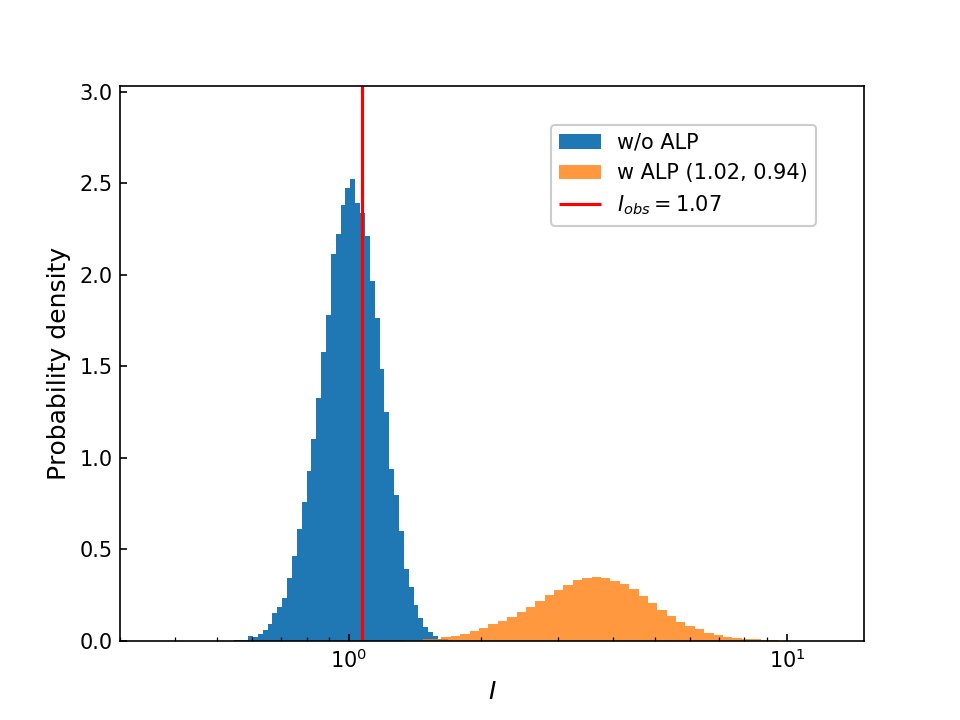}
    \caption{The probability density distribution of the spectral irregularity $I$. The blue one is the distribution for the absence of ALP effect and the orange one is under ALP parameters $m_a=1.02\;\mathrm{neV}$ and $g_{a\gamma}=0.94\;{\mathrm{GeV}}^{-1}$. The observed value $I_{\rm obs}$ is marked as red line. The $I_{\rm obs}$ is not located within the 95\% coverage of the {\it wALP} distribution, thus the corresponding ALP parameters are not favored by the observation.}
    \label{pic:hist_demo_1}
\end{figure}

ALP parameter space of $m_a$ in $(0.07-100)\;\mathrm{neV}$ and $g_{a\gamma}$ in $(0.1-7)\times 10^{-11}\;\mathrm{GeV^{-1}}$ is divided into a $(20\times 20)$ grid with logarithmic steps. For a given set of ALP parameters ($m_a$, $g_{a\gamma}$), 500 random realizations of the magnetic fields are simulated considering the random nature of the turbulent ICMF. We calculate a photon survival probability $P_{\gamma\gamma}$ for each of them. The model expected  photon counts in each energy bin thus can be calculated by
\begin{equation}
    \mu_{k}=\sum\limits_{k'}\mathcal{D}_{kk^{'}}\int\limits_{\Delta E_{k'}} P_{\gamma\gamma}(E)\frac{\mathrm{d}N}{\mathrm{d}E}(E)\mathcal{E}(E)\;\mathrm{d}E,
\label{eq12}
\end{equation}
{where $\mathcal{D}_{kk^{'}}$ is LAT energy dispersion matrix and $\mathcal{E}(E)$ is the exposure. We use the built-in commands {\tt gtexpcube2} and {\tt gtdrm}\footnote{\url{https://github.com/fermi-lat/fermitools-fhelp/blob/master/gtdrm.txt}} of {\tt Fermitools} software to obtain the $\mathcal{E}$ and $\mathcal{D}$, respectively. The integration calculation in Eq.(\ref{eq12}) is implemented using composite trapezoidal rule. Each interval of integration $\Delta{E_{k’}}$ is divided into 50 sample points (note that the interval $\Delta{E_{k’}}$ is already very small). We have tested that subdividing it into 100 points would not significantly improve the precision, since the modulated spectrum is rather smooth in a small energy interval after convolution by the instrument response function (see Fig. \ref{pic:test_the_regular_result}).} To account for the observational effect, Poisson noises are added on $\mu_{k}$ and a sample containing 100 pseudo spectra are produced for each $B$ realization and ALP parameter pair. That is, for one set of $(m_a, g_{a\gamma})$, $500\times 100$ spectra are simulated. Each one of them is fitted with a {\tt LogParabola} function and the $I$ is derived. The $I$ distribution for one $(m_a, g_{a\gamma})$ pair is presented in Fig. \ref{pic:hist_demo_1}. Our strategy of restricting ALP parameters is as follows. The null hypothesis is that the observed $I$ is drawn from the probability density function (PDF) of a given set of ALP parameters. Thus, if the observed value is located lower than 5\% of the PDF (integrate from the left side), the null can be rejected and the ALP parameters can be excluded with a confidence level of 95\%, meaning that the spectral irregularity generated by the ALP effect is too large to produce the actual observation.

\section{Results}
\label{sec:result}
Using the Eq. ({\ref{eq:m2}}), we calculate the spectrum irregularity of NGC 1275 and obtain $I_{\rm obs}=1.07$.  Such a low irregularity does not indicate existing oscillation structure in the energy spectrum.  This value is compared with the $I$ distribution of 50,000 simulations predicted by the model without ALP effect.  The resulting distribution is shown in Figure \ref{pic:hist_demo_1} as blue, which is peaked around $I\sim1$.  Fitting the distribution with a $\chi^2$ function, we can get the significance corresponding to the $I_{\rm obs}=1.07$, being 0.8$\sigma$. In order to test this result, we also use Eq. ({\ref{eq:m1}}) to calculate the irregularity, which also gives a low significance.  We therefore could conclude that there is no significant irregularity behavior in the spectrum of NGC 1275. NGC 1275 is a source with dramatic variability (see Figure \ref{pic:ngc1275_lc}), and its emissions in different periods should have diverse energy spectra. One may imagine that the superposition of the different spectra will lead to a spectral irregularity. However the results here show that it is not the case. This is because the astrophysical radiation processes always produce emission with smooth and continuous spectrum, and their superposition should also be smooth in spectrum.  Our result support that it is feasible to search for ALPs through calculating the irregularity even for the variable source. In the future work, we can further search other blazars to see if there are spectral oscillation in some sources. We can also analyze a sample of sources to explore the correlations between the $I$ and other quantities, such as strength of magnetic field, redshift, which could not only gives hints to ALP properties but also the blazar physics.

\begin{figure}
    \centering
    \includegraphics[width=0.46\textwidth]{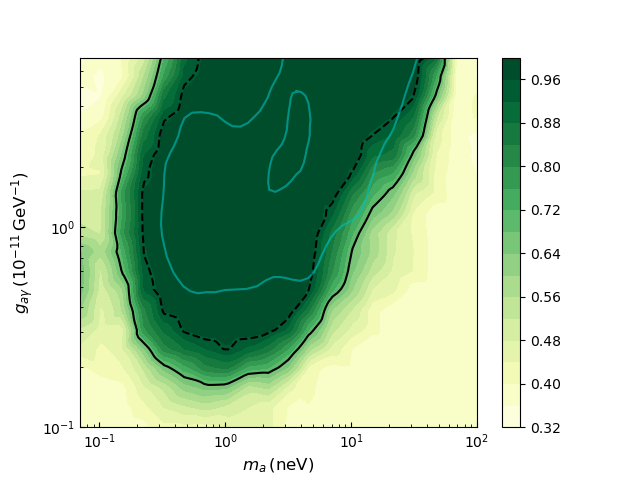}
    \caption{The excluded parameter space of ALPs based on the analysis of NGC 1275. The solid and dashed lines are the exclusion region at 68\% and 95\% confidence levels, respectively. The color represents the quantile of the distribution where the $I_{\rm obs}$ is located (start from large $I$). For a comparison, the constraints from \cite{AjelloM2016} is also plotted (cyan).}
    \label{pic:ALP_para_ngc1275}
\end{figure}

Since there is no oscillation, we constrain the ALP parameters based on our results.  As mentioned above, for each set of ALP parameters ($m_a$, $g_{a\gamma}$), we examine whether its expected $I$ distribution is consistent with the observed one. We exclude the ALP parameters with the distribution deviate from $I_{\rm obs}$ by 95\% (68\%) probability. The exclusion region is shown in Figure \ref{pic:ALP_para_ngc1275}. The solid line (dashed line) is for 95\% (68\%) C.L. As we can see from the figure, our approach can give relatively strong constraints, better than the previous works based on NGC 1275. We have ruled out $g_{a\gamma}>3\times10^{-12}\,{\rm GeV^{-1}}$ around ALP mass of $m_a\sim$ 1 neV at 95\% C.L.  In particular, the `hole-like' parameter region survived in the \cite{AjelloM2016} can be excluded here \footnote{This structure can also be excluded by the analysis of PKS 2155-304 \cite{ZhangC2018}}.

\begin{figure*}[t]
        \includegraphics[width=0.46\textwidth]{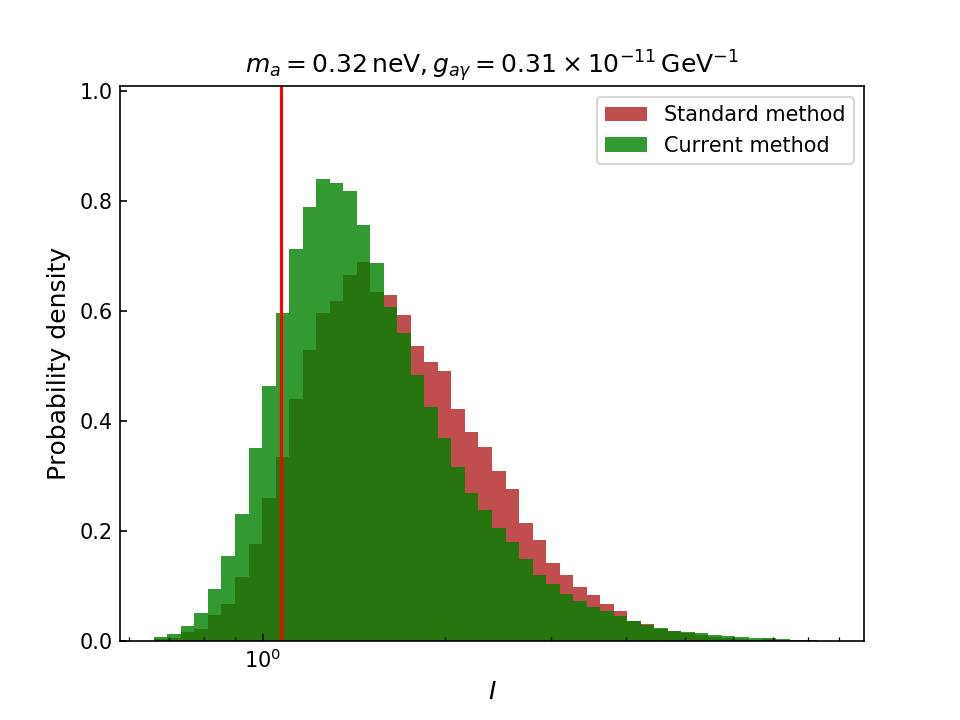}
        \includegraphics[width=0.46\textwidth]{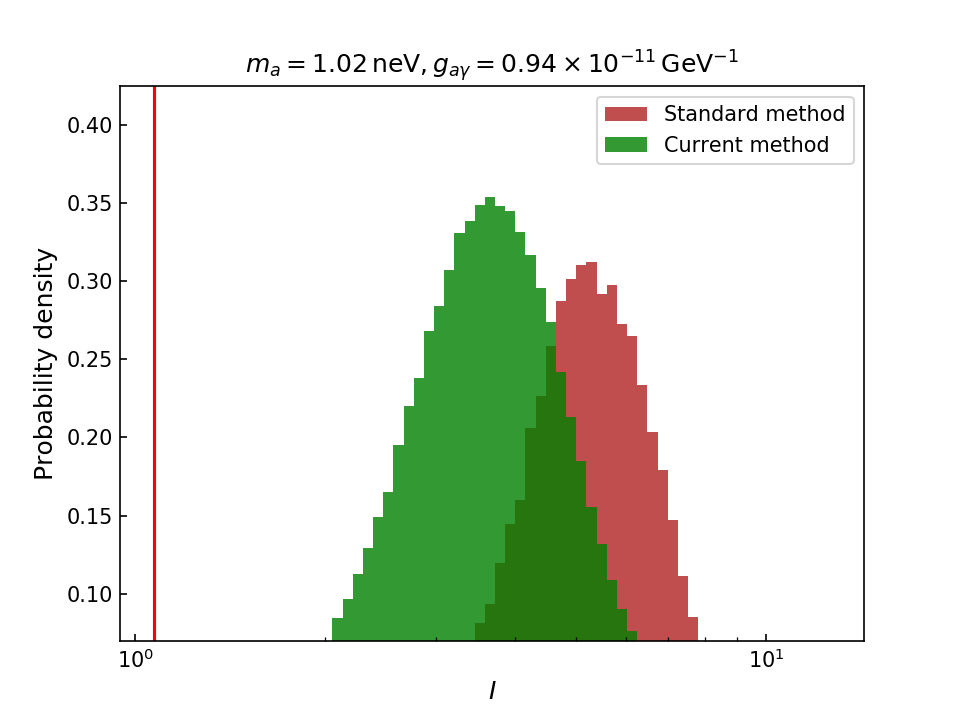}
    \caption{Comparisons of the $I$ distributions between standard method and the method taking into account all Poisson fluctuations within the ROI. The red vertical line represents for $I_{obs}=1.07$.}
    \label{pic:hist_demo_standard_current}
\end{figure*}

We need to clarify that during the simulation of giving the $I$ distribution, Poisson fluctuation is only taken into account for the target sources (NGC 1275), but not included for neighboring sources and backgrounds. We test how would this affect our results in the following. To take into account all these fluctuations, we use the {\tt Fermipy} python package to generate mock counts cubes of the ROI, and then perform likelihood analysis on the mock data. In our analysis framework, we simulate pseudo spectra under signal models, therefore for each ($m_a$, $g_{a\gamma}$) pair we need to perform the whole simulation once, which is very time-expensive. Therefore, we only choose two sets of ($m_g$, $g_{a\gamma}$) to demonstrate the influence. One is at the very center of the exclusion region and the other is close to the boundary.

Based on the best-fit model map of the ROI, we add Poisson fluctuations to it to generate pseudo counts cubes, then likelihood analysis is performed on these pseudo data. 
For each set of ALP parameters, we simulate 50,000 counts cubes (100 pseudo spectra per realization, totally 500 $B$ realizations), and obtain the corresponding SEDs. The subsequent steps are the same as the standard method. The obtained $I$ distributions are present in Fig. \ref{pic:hist_demo_standard_current}. As shown in the figure, the $I$ values by average are smaller than those from the standard method, but the difference is not great. In particular, for the set of parameters close to the boundary of the previous exclusion region (left panel of Fig. \ref{pic:hist_demo_standard_current}), the new distribution is very close to the standard one. This may be due to that the NGC1275 is a very bright gamma-ray source, so that the influence by the background fluctuation is subdominant.
Therefore, we conclude that the differences between two methods would not significantly affect the results.

\section{Discussion}
\label{sec:discussion}

\subsection{Joint constraints with PKS 2155-304}
\begin{figure*}
    \centering
    \includegraphics[width=0.46\textwidth]{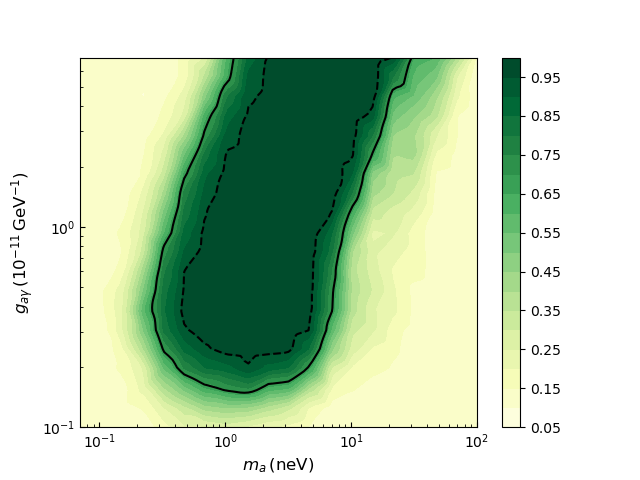}
    \includegraphics[width=0.46\textwidth]{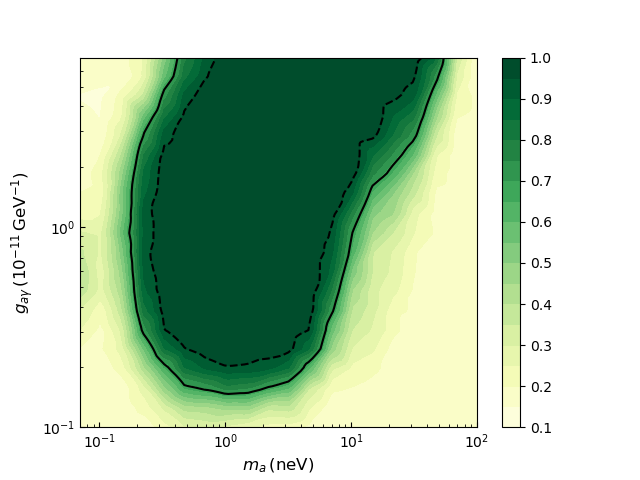}
    \caption{The excluded parameter space of ALPs based on the analysis of PKS 2155-304 ({\it left panel}) and the joint analysis of NGC 1275 and PKS 2155-304 ({\it right panel}). The solid and dashed lines are the exclusion region at 68\% and 95\% confidence levels, respectively.}
    \label{pic:ALP_para_joint}
\end{figure*}

We also analyze the source PKS 2155-304, which has been widely used to study the ALP properties.
We analyze 12-years Fermi-LAT data of PKS 2155-304 (4FGL J2158.8-3013) and constrain the ALP parameters with the same procedure as described in Sec.\ref{sec:data}.
For the ICMF parameters, we adopt the values in \cite{ZhangC2018} ($\sigma_{B}=10\;\mathrm{\mu G}, {\Lambda}_{\mathrm{min}}=0.5\;\mathrm{kpc}, {\Lambda}_{\mathrm{max}}=20\;\mathrm{kpc}, q=-11/3, \eta=0.5$). 
By calculating the irregularity of the spectrum, we obtain $I=1.2$, corresponding to a significance of 1.5 $\sigma$.
The constraints of the ALP parameter based on this source are shown in the left panel of Fig. \ref{pic:ALP_para_joint}. The results are consistent with that presented in \cite{ZhangC2018}.

One of the advantages of the approach used in this article is that we can easily combine multiple sources together to improve the constraints. Here we show the example of combining NGC 1275 and PKS 2155-304.
The combined irregularity is defined as $I_{\rm obs}=\sum\chi^2_i/{\sum d_i}$. For any given ALP parameters ($m_a$, $g_{a\gamma}$) or a w/oALP model, the distribution of $I_{\rm com}$ can be easily derived by the distribution of $I_{\rm NGC 1275}$ and $I_{\rm PKS 2155-304}$ generated before.
As a result, by comparing the observed $I_{\rm com}$ with the model expected distribution of $I_{\rm com}$, we derive the combined constraints (see right panel of Fig. \ref{pic:ALP_para_joint}). We can see that, by combining these two sources together, the constraints are further improved, better than the results reported before \citep{AjelloM2016, AdamanePallathadkaG2020}. We suggest that with more sources taken into account it will hopefully to further improve the results.

\begin{figure}[!th]
    \centering
    \includegraphics[width=0.49\textwidth]{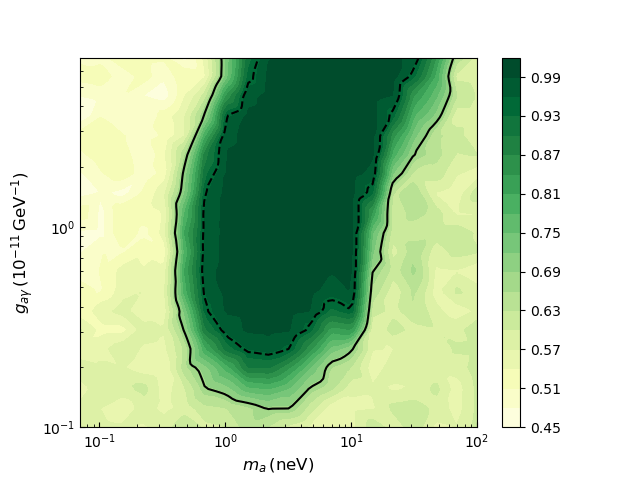}
    \caption{The projected constraints on ALP parameters with NGC 1275 assuming the energy resolution of detection instrument could be improved to $1.5\%$. Only data in the energy range 1 GeV-1 TeV are considered.}
    \label{pic:ALP_para_dampe}
\end{figure}

\subsection{Projected constraints on ALP parameters with improved energy resolution}
We would like to discuss how the energy resolution of the instrument would alter the results here. The irregular structure caused by the ALP effect would vary rapidly with energy, especially when the turbulent magnetic fields are taken into account \cite{Abramowski2013}.
Poor energy resolution will eliminate these structures, leading to a loss of the sensitivity. A high energy resolution may thus be crucial in the ALP research. At present, in the range of GeV-TeV energy band, the DArk Matter Particle Explorer (DAMPE)  \citep{dampe} has the best energy resolution, which can achieve  $<$1.5\% above $\sim1$ GeV. 
However, the effective area ($a_{\rm eff}$) of DAMPE is small. We try to use the $a_{\rm eff}$ and energy dispersion of DAMPE to perform a simulation, we find that no ALP parameters in the parameter region considered in this paper can be constrained.  Therefore, we further consider the following case. We assume that in the future, an instrument can observe the same amount of photons of NGC 1275 as Fermi-LAT in the energy range of 1 GeV-1 TeV, but has an energy resolution of 1.5\%. 
We examine what ALP parameters can be probed by such a instrument.
The simulation is similar to that in Sec. \ref{sec:data}, except we adopting an energy dispersion  of a Gaussian function with $\sigma_E=0.015 E_{\rm true}$, where $E_{\rm true}$ is the true energy of the incident photon.
We assume that no spectral irregularity is observed, namely $I_{\rm obs}\sim1$. The results are presented in Figure \ref{pic:ALP_para_dampe}. In the region of $m_a<1\;\mathrm{neV}$, the constraints are weaker than above because we only consider $> 1\;\mathrm{GeV}$ data.  
It can be seen that the exclusion region is not enlarged too much although the energy resolution has been improved significantly.  Similar results has also been demonstrated in \cite{LiangYF2019}. This results indicate that the effect of energy resolution on ALP search is not as strong as we imagined.

 \begin{figure}[th]
    \centering
    \includegraphics[width=0.49\textwidth]{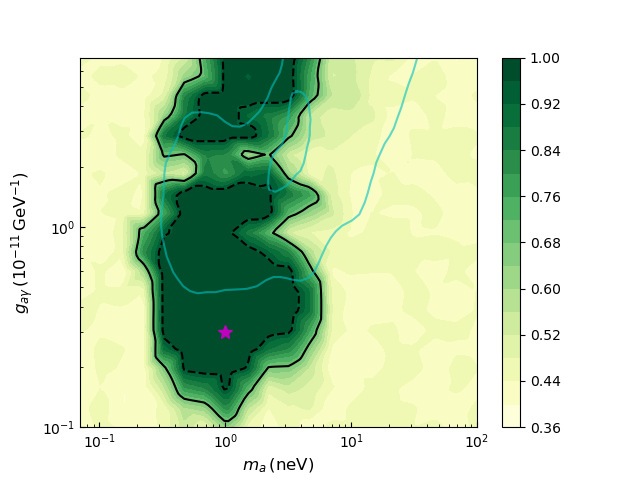}
    \caption{The excluded parameter space of ALPs based on the regular magnetic field of Perseus \cite{Libanov2020}. The solid and dashed lines are the exclusion region at 68\% and 95\% confidence levels, respectively. The color represents the quantile of the distribution where the $I_{\rm obs}$ is located (start from large $I$). For a comparison, the constraints from \cite{AjelloM2016} is also plotted (cyan). The red point corresponds to the parameters we use to examine the results (see text for details).}
    \label{pic:ALP_para_ngc1275_regularField}
\end{figure}

\begin{figure*}
    \centering
    \includegraphics[width=0.49\textwidth]{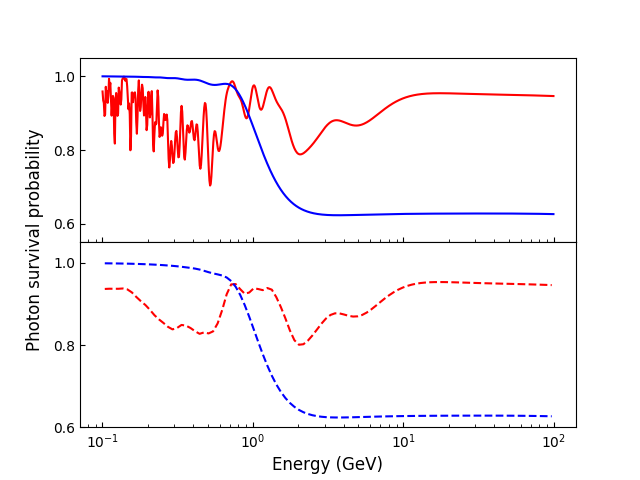}
\includegraphics[width=0.49\textwidth]{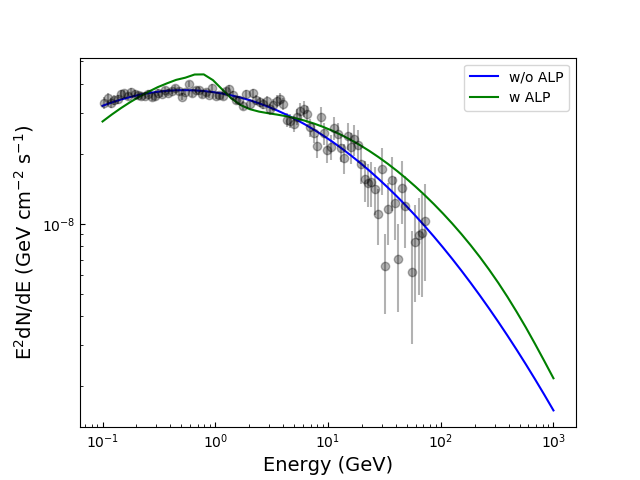}
    \caption{({\it left panel:}) The photon survival probability $P_{\gamma\gamma}$ for the representative pair of ALP parameters ($m_a=1\,{\rm neV},\,g_a=0.3\times10^{-12}\,{\rm GeV^{-1}}$). The upper panel is for the raw $P_{\gamma\gamma}$ function, while the bottom panel is the one after convolution by the Fermi-LAT energy dispersion function. The blue and red curves are for regular and turbulent ICM magnetic fields, respectively. ({\it right panel:}) For the same set of ALP parameters, best-fit ALP and w/o-ALP models are shown together with the Fermi-LAT observation.}
    \label{pic:test_the_regular_result}
\end{figure*}

\subsection{Constrains with regular magnetic field}
{The ALP-photon mixing depends largely on the morphology and strength of the magnetic field in which the $\gamma$-ray emitter resides. Following Ref.\cite{AjelloM2016}, a purely turbulent ICM is considered when deriving the above results. Ref.\cite{Libanov2020} pointed out that the use of a purely turbulent magnetic field in the Perseus cluster is not supported by any Faraday Rotation Measure observations. Therefore, they adopted a regular magnitude field to derive the constraints on ALP-photon coupling. Such a regular field is associated with the X-ray cavity near NGC 1275 and is capable of reproducing its RM observation of $6500\sim7500\, \rm{rad\,m^{-2}}$ \cite{Taylor2006}. 
It is found that the constraints are relaxed by orders of magnitude due to the reduction of spectral irregularity induced by ALP (please see \cite{Libanov2020} for details).

Here we test how would the limits change when using the regular magnetic field.
Following the magnetic field configuration described in \cite{Gourgouliatos2010} (namely the one adopted in \cite{Libanov2020}), we derive the result presented in Fig. \ref{pic:ALP_para_ngc1275_regularField}. As is shown, while the exclusion region is not the same as the turbulent one, a large amount of ALP parameters can be constrained. Furthermore, around $m_a\sim1.0\,{\rm neV}$, the 90\% upper limits of $g_{a\gamma}$ is down to $\sim10^{-12}\,{\rm GeV^{-1}}$, close to the boundary that ALP could account for all dark matter \cite{AjelloM2016}.

The result of Fig. \ref{pic:ALP_para_ngc1275_regularField} is obviously inconsistent with those reported in \cite{Libanov2020}.
To examine whether the excluded parameters are really disfavored by the observation, for a representative pair of ALP parameters (magenta point in Fig. \ref{pic:ALP_para_ngc1275_regularField}) we compare the model expected ALP spectrum with observation in Fig. \ref{pic:test_the_regular_result}. 
We find that although the regular magnetic field would not introduce dramatic oscillations in the spectrum, there is a rapid decrease of $P_{\gamma\gamma}$ around the critical energy ($\sim$GeV) followed by a constant $P_{\gamma\gamma}\sim0.6$, making the modulated spectrum deviate from the observed LogParabola-like one. For many ALP parameters, the flux varies at a level of ~30\%, greater than the error bars of the Fermi-LAT measurements. As a result, many ALP parameters can still be constrained.

Since an alternative method has been used in our analysis, one may wonder if this is the reason for the inconsistency. For the selected pair of parameters, we also use the $\chi^2$ fit to compare models. We fit the observed spectrum with both ALP and w/o-ALP model and derive their minimum $\chi^2$. The best fitted models are shown in the right panel of Fig. \ref{pic:test_the_regular_result}. We obtain a $\Delta\chi^2\sim327$. Such a large $\Delta\chi^2$ we believe can exclude the ALP model, although a null distribution based on Monte Carlo simulation is required to give precise judgement.

We note that recently Meyer et al. released their code {\it gammaALPs} for ALP calculation online \footnote{\url{https://gammaalps.readthedocs.io/en/latest/}}. To check the correctness of our program and to ensure the discrepancy is not due to errors in our calculation, we also use their code to derive corresponding results. Not strangely, since our calculation procedure mainly refers to \citet{MeyerM2014a} (and the references therein), we find that our results consistent with theirs very well. Therefore, we tend to think that the inconsistency is caused by other reasons (e.g. choice of model parameters), which need to be further investigated. The code for our calculation of the ALP-photon conversion is also put online now \footnote{\url{https://github.com/lyf222/alpconv}}.}

{\it We have confirmed that the inconsistency between the two works is mainly due to the longer data set used in our analysis. In Ref. \cite{Libanov2020}, they used 6 years of Fermi-LAT data with EDISP3 event type, while we are using 12-year data without the event-type selecion ({\tt evtype=3}). Together with the fact that NGC 1275 is in a flaring state in recent years (see fig), therefore the data amount in Ref. \cite{Libanov2020} is $\sim1/20$ of ours. The non-linear dependence of the ALP effect on the ALP parameters further makes the constraints different for $\sim$2 orders of magnitude.}

\section{Summary}
ALP-photon oscillation in external magnetic fields will induce spectral irregularity of high-energy gamma-ray sources. The degree of irregularity, $I$, can be quantified with Eq. (\ref{eq:m2}). A low observed $I_{\rm obs}$ in the spectrum of astrophysical source disfavor the ALP parameters predicting high values of irregularities. 
In this work, we revisit 12 years of Fermi-LAT observation towards NGC 1275, a source which has been widely considered in ALP studies. We calculate the spectral irregularity and then compare it to Monte Carlo simulations to test whether the corresponding ALP parameters are excluded.
We find NGC 1275 has a low degree of irregularity of $I_{\rm obs}=1.07$, indicating no significant evidence for ALPs. We thus can place constraints on ALP parameters more stringent than previous works and rule out $g_{a\gamma}>3\times10^{-12}\,{\rm GeV^{-1}}$ around ALP mass of $m_a\sim$ 1 neV at 95\% C.L. The “holelike” feature in the parameter region not probed in \cite{AjelloM2016} has also been excluded.
We show that the constraints can be further improved by combining the observation of PKS 2155-304, reaching $g_{a\gamma}\sim2\times10^{-12}\,{\rm GeV^{-1}}$ (assuming $\delta_B=10\,{\rm\mu G}$ for PKS 2155-304). It is thus worthwhile to combine more other sources in future work to enlarge the parameter region. We also find that a better instrument energy resolution would not enhance the constraints too much.

It has been pointed out that a purely turbulent magnetic field is difficult to reconcile with the rotation measure of Perseus cluster, a regular magnetic field component should be included in the ALP analysis \cite{Libanov2020}.
We therefore also derive constraints on $g_{a\gamma}$ with the purely regular field. We however obtain different results comparing to the Ref. \cite{Libanov2020}.
We find that although the regular magnetic field would not introduce dramatic irregularity in the spectrum, a prominent step around $E_{\rm c}$, which is induced by the transition to the strong mixing regime, makes the modulated spectrum not consistent with the observed LogParabola-like one. As a result, many ALP parameters can still be constrained, even down to $g_{a\gamma}\sim2\times10^{-11}\,{\rm GeV^{-1}}$ around $m_a\sim 1\,{\rm neV}$.


\section*{Acknowledgements}
We thank the anonymous referee for the evaluation of our paper. 
We thank Sergey Troitsky for his valuable discussions.
We thank Yi-Zhong Fan for the helpful suggestion. 
This work has made use of data and software provided by the Fermi Science Support Center.
This work is supported by the National Natural Science Foundation of China (Nos. 11851304, U1738136, 11533003, U1938106), the Guangxi Science Foundation (2017AD22006, 2018GXNSFDA281033), special funding for Guangxi distinguished professors and Innovation Project of Guangxi Graduate Education (YCBZ2021025).

\bibliographystyle{apsrev4-1-lyf}
\bibliography{references}

\end{document}